\documentclass{osa-article}

\journal{oe}

\DeclareMathOperator{\sech}{sech}

\begin{document}

\title{Supersensitive estimation of the coupling rate in cavity optomechanics with an impurity-doped Bose-Einstein condensate}
\author{Qing-Shou Tan,\authormark{1,2*} Ji-Bing Yuan,\authormark{3$\dagger$} Jie-Qiao Liao,\authormark{4$\ddagger$} Le-Man Kuang \authormark{4$\S$}}
\address{
\authormark{1}Key Laboratory of Hunan Province on Information Photonics and Freespace Optical Communications, Hunan Institute of Science and Technology, Yueyang 414000, China\\
\authormark{2}College of Physics and Electronic Engineering, Hainan Normal University, Haikou 571158, China\\
\authormark{3}Department of Physics and Electronic Information Science, Hengyang Normal University, Hengyang 421002, China\\
\authormark{4}Key Laboratory of Low-Dimensional Quantum Structures and Quantum Control of Ministry of Education,\\
Department of Physics and Synergetic Innovation Center for Quantum Effects and Applications, Hunan Normal University, Changsha 410081, China}

\email{\authormark{*}qingshoutan@163.com }
\email{\authormark{$\dagger$}yuanjibing1@163.com }
\email{\authormark{$\ddagger$}jqliao@hunnu.edu.cn} 
\email{\authormark{$\S$} lmkuang@hunnu.edu.cn}

\begin{abstract}
We propose a scheme to implement a supersensitive estimation of the coupling strength in a hybrid optomechanical system
which consists of a cavity-Bose-Einstein condensate system coupled to an impurity. This method can dramatically improve the estimation precision even when the involved photon number is small. The quantum Fisher information indicates that the Heisenberg scale sensitivity
of the coupling rate could be obtained when the photon loss rate is smaller than the corresponding critical value in the input of either coherent state or squeezed state. The critical photon decay rate for the coherent state is larger than that of the squeezed state, and the coherent state input case is more robust against the photon loss than the squeezed state case.
We also present the measurement scheme which can saturate the quantum Cram\'er-Rao bound.
\end{abstract}

\section{Introduction \label{introduction}}

Cavity optomechanics studies the radiation pressure coupling between the optical modes and the mechanical oscillation~\cite{dorsel,ritsch}. Cavity optomechanical systems intrinsically integrate the advantages of precision detection (optical mode) and signal sensing (mechanical oscillation), and thus the optomechanical systems provide a useful platform for the study of high-precision metrology~\cite{tenfel,zhang,aspelmeyer, Pezze,liao,liao2,liao3}. Recently, a lot of investigation has been devoted to the estimation of physical quantities based on various optomechanical systems~\cite{Rae,Restrepo,Zhou,Zhangk,Jingh,Brahms,Kohler,li}. These measurement schemes include weak force detection~\cite{clerk,tsang,wimmer,Zhao}, rotation sensing~\cite{davuluri,davuluri2}, mass
sensing~\cite{yang,lin}, gravity estimation~\cite{Armata,Qvarfort}, and gravitational wave detection~\cite{Pace}.

The cavity optomechanical systems can be implemented with various physical setups. In particular, the system of ultracold atoms has  some advantage over other systems to realize the optomechanical interaction: (i) The thermal noise in this system can be largely suppressed because the thermal effect in ultracold atoms is negligible. (ii) The strong coupling regime at the level of single quanta can be realized owing to the collective enhancement effect in the interaction between the cavity field and the atoms. (iii) There are good and tested experimental techniques in the fields of atomic Bose-Einstein condensates (BECs) and cavity quantum electrodynamics~\cite{Pinkse,Brennecke,Singh,Mekhov,Zhang,chen,Maschler,Yang}. These features motivate various applications based on the utilizing of genuine quantum optomechanical effects. Recently, some  novel results enabled by BEC optomechanics  have also been reported, such as squeezed rotors~\cite{Buchmann}, super-radiance phonon laser~\cite{Jiang} and  spin-orbit effects~\cite{Deng}.

In this paper, we propose to study the supersensitive estimation of the coupling parameters related to the optical and oscillation degrees of freedom in a hybrid optomechanical system. This system is formed by an optical cavity, a cigar-shaped BEC, and a two-level impurity atom immersed in the BEC. Here, both the BEC and the impurity atom are coupled to the cavity
field in the dispersive regime. The coupling between the BEC atoms and the cavity mode takes a form as the optomechanical-type interaction, and the coupling between the two-level impurity and the BEC is described by dispersive (diagonal) interaction~\cite{cirone,haikka,yuan,tan,song}. We consider the optical cavity in either a coherent state or a squeezed vacuum state.
To estimate the coupling rate, we calculate the quantum Fisher information (QFI)~\cite{ braunstein,Salasnich,paris,jing, ymzhang,liu,liu2,liu3,nori} of the
impurity part in the final state, which is related to the quantum Cram\'er-Rao (QCR)
bound on the precision of estimator. We find that the injection of either coherent or squeezed light into the
optomechanical cavity can  improve the estimation precision to
the Heisenberg limit (HL)  when the photon loss rate is smaller than the critical values.
The squeezed vacuum state has advantage over that of the coherent state~\cite{Zhao} under small photon decay rate. As a tradeoff, however, the
squeezed state scheme is fragile to the photon loss. This is because the squeezing scheme is not only sensitive to the change of the estimated parameters, but also to the noise. In this sense, the coherent state is more robust against the
photon loss of the cavity.

The rest of this work is organized as follows. In Sec.~\ref{model}, we introduce the physical model and present the Hamiltonian of the system. In Secs.~\ref{Dynamicsatom} and~\ref{Finformation}, we calculate the dynamics of the impurity atom and the Fisher information corresponding to the coherent state and squeezed state inputs of the cavity field, respectively. In Sec.~\ref{couprestimaiton}, we study how to realize the supersensitive estimation of the coupling parameters. A conclusion will be presented in Sec.~\ref{conclusion}. Finally, we add an appendix to show a detailed derivation of the effective Hamiltonian in the large detuning case.

\section{Model and Hamiltonian \label{model}}
\begin{figure}[ht]
\centering\includegraphics[scale=0.66]{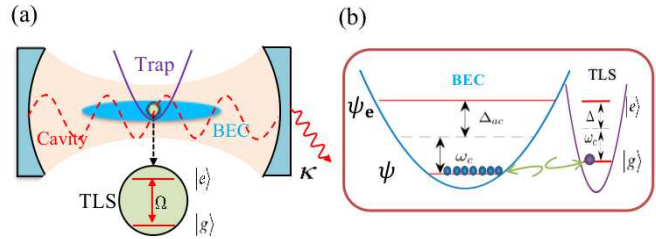}
\caption{(a) Scheme of the hybrid cavity optomechanical system for coupling
strength estimation. The cavity confines the photons dispersively coupled to
a BEC. An impurity atom, described as a two-level system (TLS) $|e\rangle$ and $|g\rangle$, is immersed
in the Bose gas. (b) Both the  BEC and  impurity atoms are coupled to the cavity field via the dispersive coupling,
$\Delta_{ac}$ and $\Delta$ are detunings.  Besides, the 
BEC atoms also couple with the ground state $|g\rangle$ of the impurity atom via $s$-wave scattering interaction.  }
\label{setup}
\end{figure}
We consider a hybrid optomechanical system which is formed by a
cigar-shaped BEC of $N$ two-level atoms placed in a cavity  and
a two-level impurity atom immersed inside the BEC  (as depicted in Fig.~\ref{setup}).
We study the case where the BEC atoms are coupled to the cavity field via the dispersive coupling,
 the coupling between these atoms are described by the $s$-wave scattering interaction.
 For the case of large detuning, the atomic transition to the excited states are suppressed.  In the frame rotating at the cavity resonance frequency, we can obtain the effective Hamiltonian of the hybrid system  as follows (see the Appendix for details): 
\begin{eqnarray}\label{h1}
H_{\rm eff}=H_1+H_2+H_3.
\end{eqnarray}
Below, we will introduce these three parts in detail.

In this system, the coupling between the cavity field and the BEC can be described by the Hamiltonian~\cite{Brennecke,Singh}
\begin{eqnarray}
\label{h11}
H_{1}=\int dx{\Psi}^{\dag}(x)\left[\frac{-\hbar^{2}}{2m}
\frac{d^{2}}{dx^{2}}+V_{\rm ext}(x)+\hbar U_0\cos^{2}(kx)a^{\dag}a +\frac{g_{1D}}{2}|\Psi(x)|^{2}\right]{\Psi}(x),
\end{eqnarray}
where $a^{\dag}$ and $a$ are the creation and annihilation operators of the cavity field with the resonance frequency $\omega_{c}$, $\Psi^{\dag}(x)$ and $\Psi(x)$ are the creation and annihilation operators of the Schr\"odinger field describing the atomic BEC, $m$ is the mass of a single atom.  $V_{\rm ext}(x)$  is the weak external trapping potential.
In the dispersive coupling region,  the BEC atom-cavity photon interaction induces an additional potential $U_0\cos ^{2}(kx)a^{\dagger}a$ for atoms with the wave vector $k$.
Here $U_{0}$ $=-g^{2}/\Delta_{ac}$ is the maximal light shift per photon that an
atom may experience, with $g$ being the atom-photon coupling constant and $\Delta_{ac}=\omega_{a}-\omega_{c}$ the detuning between the BEC atomic resonance frequency $\omega_a$ and cavity frequency $\omega_c$.   Besides, $g_{1D}$ is the atom-atom interaction strength integrated along the transverse directions.
 
In the low-temperature limit, we assume that the impurity atom is in the
ground state of harmonic trap $V(x)=m_{A}\omega_{A}^{2}x^{2}/2$ that is independent of the internal states,
and hence the spatial wave function is
$\varphi (x)={\pi ^{-1/4}\ell ^{-1/2}}\exp (-x^{2}/2\ell ^{2})$,
where $\ell=\sqrt{\hbar/(m_{A}\omega_{A})}$ with $\omega_A$  being the trap
frequency and $m_{A}$ the mass of the impurity.
Assuming  that the impurity atom undergoes $s$-wave collisions with
the BEC atoms only when the impurity atom is in its  ground state, and the
corresponding scattering length is $a_{AB}$~\cite{haikka}. 
Then the impurity-BEC
interaction can  be described by the following Hamiltonian~\cite{yuan}
\begin{eqnarray}
H_{2}=\frac{4\hbar^2 a_{AB}}{m_{AB}(\ell_{A\bot}^{2}+\ell_{B\bot}^{2})}|g\rangle\langle g|\int dx|\varphi (x)|^{2}\Psi^{\dag}(x)\Psi(x), \label{ham7} 
\end{eqnarray}
 with $m_{AB}=m_{A}m/(m_{A}+m)$ being the reduced mass, $\ell _{A\bot }$ and $ \ell _{B\bot }$ are the harmonic oscillator lengths of
the impurity and BEC in the transverse directions.

Finally,  after adiabatically eliminating the  coherent transition between the ground state and the excited state, the  interaction between the cavity and the impurity can be obtained  as 
\begin{eqnarray}
\label{ham9}
H_{3}=\hbar \Delta \vert e\rangle \langle
e\vert +\hbar\delta_{ac}\sigma_{a}^{z}(2 a^{\dag }a+1),
\end{eqnarray}
with the  Pauli operator $\sigma_{a}^z=|e\rangle\langle e|-|g\rangle\langle g|$ and $\Delta=\Omega-\omega_{c}$ being the detuning between the cavity mode and the impurity resonance frequency.
Here, $\hbar \Omega$ is level splitting between the ground state $|g\rangle$ and excited  state $|e\rangle$, 
$\delta _{ac}={g_{ac}^{2}}/{\Delta}$
 and $g_{ac}=g_{0}\int
dx\varphi (x)\cos (kx)$ with $g_{0}$ being the maximum impurity-cavity
coupling strength.

As we are interested in the  case of weak atom-atom interactions $g_{1D}\sim0$, which  is tunable through the Feshbach resonance, and therefore the macroscopically occupied zero momentum state is coupled to the symmetric superposition of the $\vert\pm 2\hbar k\rangle$ momentum states via absorption and stimulated
emission of cavity photons~\cite{Brennecke,Singh}. The corresponding Schr\"odinger field can be expanded as the
following single-mode quantum field
\begin{eqnarray}\label{psix}
\Psi(x)= b_{0}\sqrt{1/L}+b\cos(2kx)\sqrt{2/L}
\end{eqnarray}
with $L$ being the length of the BEC which has been extended
symmetrically about the center of the cavity.
Here $b$ is the annihilation operator
of Bogoliubov mode corresponding to the quantum fluctuation of the atomic
field about the condensate mode $b_{0}$.
Assuming that depletion of the zero-momentum component of the condensate is small,
we adopt the classical treatment via the replacement $b_{0}^{\dag },b_{0}\rightarrow \sqrt{N}$.
Substituting Eq.~(\ref{psix}) into Hamiltonian (\ref{h1}) and neglecting unimportant constant terms, then
the effective Hamiltonian of the
hybrid system can be written as 
\begin{align}\label{ham2}
H_{\rm eff}'  = & \hspace{0.1cm} \hbar\omega _{m}b^{\dag }b+\hbar(\Omega _{c}-i\kappa )a^{\dag}a 
+\hbar(\Omega _{A}-i\gamma)\left\vert e\right\rangle \left\langle e\right\vert +\hbar G\left\vert e\right\rangle \left\langle e\right\vert a^{\dag }a\nonumber\\
& +\hbar(\chi a^{\dag }a+g_{ab}\left\vert g\right\rangle \left\langle g\right\vert )(b+b^{\dag }),
\end{align}
where $\omega_{m}=\hbar(2k)^{2}/(2m)=4\omega_{\mathrm{rec}}$ with $\omega_{\mathrm{rec}}$ being the recoil frequency of the condensate atoms.
In Eq.~(\ref{ham2}), $\Omega_{c}=U_0N/2-2\delta_{ac} $ and $\Omega
_{A}=\Delta+2\delta_{ac}-\delta_{ab}$ denote the
effective oscillation frequency for cavity and impurity, respectively,  with $\delta_{ab}=4N\hbar a_{AB}/[m_{AB}(\ell _{A\bot }^{2}+\ell _{B\bot }^{2})L] $
being the level shift due to the collision between the impurity and BEC.
Besides, $G=4\delta_{ac}$ is the coupling strength between cavity and impurity. 
The last term in Eq.~(\ref{ham2})  involves  an optomechanical interaction
with the coupling strength  $\chi=U_{0}\sqrt{N/8}$, and $g_{ab}={4\sqrt{2 N}\hbar a_{AB}e^{-\ell ^{2}k^{2}}}[{m_{AB}(\ell
_{A\bot }^{2}+\ell _{B\bot }^{2})L}]^{-1}$ is the impurity-BEC coupling parameter.
In Eq.~(\ref{ham2}), we phenomenally introduce the cavity photon loss $\kappa $  and the impurity atom spontaneous decay rate $\gamma$,  due to the  interactions  with the vacuum environment.

\section{ Dynamics of the  impurity atom \label{Dynamicsatom}}
In the interaction picture with respect to  $\omega_m b^{\dagger}b$,  the Hamiltonian $H_{\rm eff}'$ becomes
\begin{align}
H_I(t)=& \hspace{0.1cm}(\Omega _{c}-i\kappa )a^{\dag
}a + (G a^{\dag }a+\Omega _{A}-i\gamma)\left\vert e\right\rangle \left\langle e\right\vert \nonumber \\
&+(\chi a^{\dag }a+g_{ab}\left\vert g\right\rangle \left\langle
g\right\vert )[b \exp{(-i\omega_m t)}+b^{\dagger}\exp({i\omega_m t})].
\end{align}
The time evolution operator associated with $H_I(t)$ can be obtained by using the Magnus expansion as~\cite{Blanes,bar-gill} (hereafter we set $\hbar=1$)
\begin{eqnarray}
U(t)\equiv {\mathcal T}_{+} \exp\left[-i \int_0^{t} H_I(t')dt' \right]=\exp\left[\sum_{n=1}^{\infty}\frac{(-i)^n}{n!}F_n(t)\right],
\end{eqnarray}
where  ${\mathcal T}_{+} $ is  the time-ordering operator.

Note that in our case only the first two terms of  $F_n(t)$ are nonzero:
\begin{align}
F_1(t) \equiv &\int_0^{t} H_I(t')dt'  =  \left[(\Omega_{c}-i\kappa)a^{\dagger}a+(G a^{\dagger}a+\Omega_{A}-i\gamma)\left\vert e\right\rangle \left\langle e\right\vert \right]t  \nonumber\\
 & \hspace{2.4cm}+ i ({\chi} a^{\dagger}a+{g}_{ab}\left\vert g\right\rangle \left\langle g\right\vert ) \left ( \frac{\lambda b^{\dagger}-\lambda^{\ast}b}{\omega_m} \right), \nonumber\\
 F_2(t) \equiv &\int_0^{t}ds \int_0^{s}ds' [H_I(s),H_I(s')]=-2i(\chi a^{\dag }a+g_{ab}\left\vert g\right\rangle \left\langle g\right\vert )^2 \left( \frac{\omega_m t -\sin(\omega_m t)}{\omega_m^2}\right)
\end{align}
with  $\lambda = 1-e^{i\omega _{m}t}$. This is because $[H_I(s),H_I(s')]= -2i(\chi a^{\dag }a+g_{ab}\left\vert g\right\rangle \left\langle
g\right\vert )^2 \sin[\omega_m(s-s')]$ which  commutes with $H_I(t)$.
Due to the commutor $[F_1(t), F_2(t)]=0$,  the time evolution operator governed by Hamiltonian~(\ref{ham2}) can be obtained as
\begin{align}\label{ut}
U(t)  =& \exp\left[ -i F_1(t) -\frac{1}{2}F_2(t)\right]\nonumber\\
=&  \exp\left\{-i \left[\Omega_{c}a^{\dagger}a+(G a^{\dagger}a +\Omega_{A})\left\vert e\right\rangle \left\langle e\right\vert \right]t \right \} \exp[(\tilde{\chi} a^{\dagger}a+\tilde{g}_{ab}\left\vert g\right\rangle \left\langle g\right\vert )(\lambda b^{\dagger}-\lambda^{\ast}b)]\nonumber\\
 & \times\exp\left\{ i(\tilde{\chi} a^{\dagger}a+\tilde{g}_{ab}\left\vert g\right\rangle \left\langle g\right\vert )^{2}[\omega_{m}t-\sin(\omega_{m}t)]\right\} \exp(-\kappa a^{\dagger}at)\exp(-\gamma\left|e\right\rangle \left\langle e\right|t),
\end{align}
where  $\tilde{\chi}\equiv\chi /\omega _{m}$ and $\tilde{g}_{ab} \equiv
g_{ab}/\omega _{m}$ are rescaled dimensionless coupling strengths.

To proceed, we consider  the initial state of the total system as
\begin{eqnarray}  \label{rou}
\rho ^{T}(0)=\vert\phi\rangle\langle\phi\vert \otimes \rho^{B}\otimes\vert\psi\rangle_{\rm imp}\langle\psi\vert,
\end{eqnarray}
where $\vert \phi \rangle $ is an arbitrary input
states of the cavity photon and $\rho ^{B}$ is the thermal equilibrium state of the
Bogoliubov mode, defined by $\rho ^{B}=[1-\exp(-\beta\omega_{m})]\exp(-\beta\omega_{m}b^{\dag }b)$
with $\beta$ the inverse temperature. 
The impurity atom is prepared in $\vert\psi\rangle_{\rm imp}=c_e(0) |e\rangle+c_g(0) |g\rangle$
involving two-level states $|e\rangle$ and $|g\rangle$. The superposition coefficients $c_{e,g}(0)$ satisfy the normalization condition $\vert c_{e}(0)\vert ^{2}+\vert c_{g}(0)\vert ^{2}=1$.

  In terms of Eqs.~(\ref{ut}) and (\ref{rou}), we can get the final state of the total system at time $t$.  
  After tracing over the Bogoliubov mode, we can obtain
the reduced matrix elements of the impurity-cavity system,
\begin{align}\label{rou2}
\rho_{ee,mn}(t) =  & \left\vert c_{e}(0)\right\vert ^{2}\exp\left[-\kappa(m+n)t\right]\exp(-2\gamma t)\exp\left[-i\left(\Omega_{c}+G\right)(m-n)t\right]\nonumber \\
  & \times\exp\{ i[\tilde{\chi}^{2}(m^{2}-n^{2})][\omega_{m}t-\sin(\omega_{m}t)]\} \nonumber \\
 & \times\exp\left\{ - \left\vert \lambda\right\vert ^{2} \tilde{\chi}^{2}(m-n)^{2}\coth\left(\frac{\beta\omega_{m}}{2}\right)\left[1-\cos(\omega_{m}t)\right]\right\}\langle m |\phi\rangle \langle \phi| n\rangle, \nonumber\\
\rho_{gg,mn}(t)  =  & \vert c_{g}(0)\vert ^{2}\exp\left[-\kappa(m+n)t\right]\exp\left[-i\Omega_{c}(m-n)t\right]  \nonumber \\
  & \times\exp\{i[\tilde{\chi}^{2}(m^{2}-n^{2})+2\tilde{\chi}\tilde{g}_{ab}(m-n)][\omega_{m}t-\sin(\omega_{m}t)]\}\nonumber \\
  & \times \exp\left\{ - \left\vert \lambda\right\vert ^{2} \tilde{\chi}^{2} (m-n)^{2}\coth\left(\frac{\beta\omega_{m}}{2}\right)\left[1-\cos(\omega_{m}t)\right]\right\}\langle m |\phi\rangle \langle \phi| n\rangle, \nonumber\\
\rho_{eg,mn}(t)  =  & c_{e}(0)c_{g}^{\ast}(0)\exp\left[-\kappa(m+n)t\right]\exp(-\gamma t)\exp\{-i[\Omega_{c}(m-n)+\Omega_{A}+Gm]t\}\nonumber \\
  & \times\exp\{i [\tilde{\chi}^{2}(m^{2}-n^{2})-2\tilde{\chi}\tilde{g}_{ab}n-\tilde{g}_{ab}^{2}] [\omega_{m}t-\sin(\omega_{m}t)]\}\nonumber \\
  & \times\exp\left\{ - \left\vert \lambda\right\vert ^{2}  [\tilde{\chi} (m-n) -\tilde{g}_{ab} ]^2\coth\left(\frac{\beta\omega_{m}}{2}\right)\left[1-\cos(\omega_{m}t)\right]\right\}\langle m |\phi\rangle \langle \phi| n\rangle,
\end{align}
with $ \rho_{ge,mn}=\rho_{eg,nm}^{\ast }$ and $\langle n|\phi\rangle$ being the overlap between $|\phi\rangle$ and number state $|n\rangle$.

The estimation of the coupling strengths $\chi$ and $g$, equivalent to $\tilde{\chi}$ and $\tilde{g}$  when $\omega_m$ is known,
can be realized by  implementing  the estimation in the internal state of the impurity atom.
At time $t=\tau \equiv 2\pi /\omega _{m}$, after tracing out the cavity mode, the final
matrix elements of the impurity atom become
\begin{eqnarray}\label{reg}
&&\rho _{eg}(\tau ) = c_{e}(\tau)c_{g}^{\ast }(\tau)\exp \left[-i2\pi (\tilde{g}_{ab}^{2}+\tilde{\Omega}_{A}) \right]\sum_{n=0}^{\infty }\exp[ -i2\pi (\tilde{G}+2\tilde{\chi}\tilde{g}_{ab}-2i\tilde{\kappa})n]|\langle
n|\phi \rangle |^{2},  \nonumber\\
&&\rho _{ee}(\tau )= \left\vert c_{e}(\tau)\right\vert
^{2}\sum_{n=0}^{\infty }\exp( -4\pi \tilde{\kappa}n) |\langle
n|\phi \rangle |^{2},\nonumber\\
&&\rho _{gg}(\tau )=\left\vert c_{g}(\tau)\right\vert
^{2}\sum_{n=0}^{\infty }\exp( -4\pi \tilde{\kappa}n) |\langle
n|\phi \rangle |^{2},
\end{eqnarray}
where $c_e(\tau)\equiv\exp(-2\pi \tilde{\gamma})c_e(0)$ and $c_g(\tau)=c_g(0)$. Equation.~(\ref{reg}) indicate that 
the spontaneous decay $\gamma$ will
change the atomic population of the impurity atom.
Here,  the rescaled dimensionless parameters  are defined as $\tilde{x}\equiv x/\omega _{m} $ $(x\in \{ \Omega_{A}, G, \kappa, \gamma \})$.
 Based on Eq.~(\ref{reg}), we can find that $\rho_{eg}$ carries the
information of the coupling strength $\tilde{\chi}$ and $\tilde{g}_{ab}$
via the relation $O\equiv \tilde{\chi}\tilde{g}_{ab}$. 
While the diagonal elements of $\rho$ do not 
include the coupling between the atom and the cavity, due to the  dispersive (diagonal) interaction~\cite{yuan}.
In what follows,
we directly consider the precision of $O$ for estimating
the coupling strengths. The
estimation of parameter $\tilde{\chi}$ can be obtained  by
performing a homodyne measurement on the cavity field~\cite{Armata, Qvarfort}.

\section{Quantum Fisher information for coherent state and squeezed vacuum state inputs \label{Finformation}}
We  now use the impurity atom as a probe qubit to detect the  coupling rate in our hybrid system.
To obtain the best sensitivity of the estimated parameter, we introduce the QFI which gives a theoretical-achievable limit
on the precision of the estimator  $O $ via QCR  bound,
$\delta O \ge  \delta O_{\rm QCR} \equiv 1/\sqrt{s F (O)}$ with $s $ being the
number of repeated measurements (we will choose $s=1$  throughout this paper). To evaluate the QFI, we should first
normalize and diagonalize $\rho (\tau )$ as $\rho =\sum_{i}\lambda
_{i}|\varphi _{i}\rangle \langle \varphi _{i}|$ with $\sum_{i}\lambda _{i}=1$.
 Without loss of generality,  we  choose  $c_{e}(\tau)=c_{g}(\tau)$ at the working point $\tau$  to eliminate the effects of the impurity atom 
decay on QFI. 
 Then  the corresponding eigenvalues and eigenvectors of the impurity atom are given as
\begin{eqnarray}\label{eigva}
&&\lambda _{1}=\frac{1}{2}+\frac{|\rho _{eg}|}{2\rho _{ee}},\hspace{1.5cm} |\varphi _{1}\rangle =\frac{1}{\sqrt{2}}\left[ \left( \frac{\rho _{eg}}{
|\rho _{eg}|}\right) |e\rangle +|g\rangle \right], \nonumber\\
&&\lambda _{2}=\frac{1}{2}-\frac{|\rho _{eg}|}{2\rho _{ee}},  \hspace{1.5cm}
|\varphi _{2}\rangle =\frac{1}{\sqrt{2}}\left[ \left( \frac{\rho _{eg}}{
|\rho _{eg}|}\right) |e\rangle -|g\rangle \right] .
\end{eqnarray}
 Therefore, the QFI with respect to $O$ can be explicitly expressed as~\cite
{paris,jing,liu,liu2,liu3,ymzhang,nori}
\begin{eqnarray} \label{fish}
F(O)=\sum_{i=1}^{M}\frac{\left( \partial _{O}\lambda _{i}\right) ^{2}}{
\lambda _{i}}+4\sum_{i=1}^{M}\lambda _{i}\langle \partial _{O}\varphi
_{i}|\partial _{O}\varphi _{i}\rangle -\sum_{i,j=1}^{M}\frac{8\lambda _{i}\lambda _{j}}{\lambda _{i}+\lambda _{j}
}\left\vert \langle \varphi _{i}|\partial _{O}\varphi _{j}\rangle
\right\vert^{2},
\end{eqnarray}
where  $M$  denotes the number of nonzero eigenvalues $\lambda_i$.
For two-level system we have $M=2$ and $\lambda _{1}+\lambda _{2}=1$, then Eq.~(\ref{fish}) will be  further simplified   as
 $F(O)=F_{c}+F_{q}$, which contains the following two terms:
\begin{eqnarray} \label{fish3}
&&F_{c} \equiv \frac{\left( \partial_{O}\lambda_{1}\right)^{2}}{\lambda
_{1}(1-\lambda_{1})}=\frac{(\text{Re}[\mathcal{A}])^{2}|\rho_{eg}|^{2}}{\rho_{ee}^{2}-|\rho _{eg}|^{2}}, \nonumber\\
&&F_{q}\equiv 4(1-2\lambda_{1})^{2}\left\vert\langle \partial_{O}\lambda
_{1}|\lambda_{2}\rangle \right\vert^{2}=\frac{|\rho_{eg}|^{2}(\text{Im}[
\mathcal{A}])^{2}}{\rho_{ee}^{2}}.
\end{eqnarray}
Here $\text{Re}[\mathcal{A}]$ ($\text{Im}[\mathcal{A}]$) is the real
(imaginary) part of $\mathcal{A}$, defined as
$\mathcal{A}\equiv ({\partial_{O}\rho_{eg}})/{\rho_{eg}}$.
In Eq.~(\ref{fish3}), $F_c$ is regarded as the classical
contribution, whereas $F_q$ contains the truly quantum
contribution~\cite{paris}. 
In terms of the above equations, we find that the QFI relies on the values of
 $\rho _{ee},|\rho _{eg}|$ and $\mathcal{A}$ which depend on the form of
input optical states $|\phi \rangle $. Below we will consider two common
input states: the coherent state $|\alpha \rangle $ and the squeezed vacuum
state $|\xi \rangle $.

\emph{Coherent state  $|\protect\alpha\rangle$} case \textemdash
Suppose that the cavity photons are in a coherent state~\cite{Scully}
\begin{eqnarray}
\left\vert \phi \right\rangle \equiv \left\vert \alpha \right\rangle
=\exp({-\left\vert \alpha \right\vert^{2}/2})\sum_{n=0}\frac{\alpha^{n}}{\sqrt{
n!}}\left\vert n\right\rangle,
\end{eqnarray}
and by substituting  it into Eq.~(\ref{reg}), we get the corresponding matrix
elements as
\begin{eqnarray}
&&\rho_{eg}(\tau ) =\rho_{ge}^{*}(\tau ) =\frac{1}{2}{\exp\left[{-i2\pi \left( \tilde{\Omega}_{A}+\tilde{g}_{ab}^{2}\right)
}\right]\exp \left\{ \bar{n}_{|\alpha\rangle }\left[e^{-4\pi \tilde{\kappa}}e^{-i2\pi(\tilde{G}+2O)}-1\right]\right\}}, \notag\\
&&\rho_{ee}(\tau ) =\rho _{gg}(\tau )=\frac{1}{2}\exp\left[\bar{n}_{|\alpha
\rangle }(e^{-4\pi \tilde{\kappa}}-1)\right],
\end{eqnarray}
where  $\bar{n}_{|\alpha \rangle}=\left\vert \alpha \right\vert ^{2}$ is  the average photon number.

 In terms  of $\rho_{eg}$,  the expression of $\mathcal{A}$ can be
written as
\begin{eqnarray}
\mathcal{A}_{\left\vert \alpha \right\rangle }=-i4\pi \bar{n}_{|\alpha\rangle
}\exp({-4\pi \tilde{\kappa}})\exp[{-i2\pi (\tilde{G}+2O)}].
\end{eqnarray}
Based on Eqs.~(\ref{fish}) and (\ref{fish3}), we can obtain the QFI for the case of coherent state input (see Tab.~1).

\emph{Squeezed vacuum state $|\protect\xi\rangle$} case \textemdash
We now turn to the case of squeezed vacuum state input. The definition of squeezed
vacuum state is~\cite{Scully}
\begin{eqnarray}
|\phi \rangle \equiv |\xi \rangle
=\sqrt{\sech r}\sum_{n=0}^{\infty}\frac{\sqrt{(2n)!}}{2^n n!}\left[-\exp({i\vartheta})\tanh r\right]^{n}\left\vert 2n\right\rangle
\end{eqnarray}
with $r$ being the squeezing  modulus and $\vartheta$ the phase. Note that $|\xi\rangle$ is a superposition
only of even photon number states, thus we have
$\langle 2m|\xi \rangle =\frac{\sqrt{\sech r (2m)!}}{2^m m!}\left[-\exp({-i\vartheta })\tanh r\right]^{m}$.
 Substitution of Eq.~(20)  into Eq.~(\ref{reg}) leads to
\begin{eqnarray}
&&\rho _{eg}(\tau )=\rho _{eg}^{\ast }(\tau ) = \frac{\exp[{-i2\pi (\tilde{\Omega}_{A}+\tilde{g}_{ab}^{2})} ]
\sech r}{2\sqrt{1-\exp({-8\pi \tilde{\kappa}})\exp[{ -i4\pi ( \tilde{G}
+2O) }] \tanh ^{2} r}},\notag  \label{rs} \\
&&\rho _{ee}(\tau ) =\rho _{gg}(\tau )=\frac{\sech r}{2\sqrt{1-\exp({-8\pi
\tilde{\kappa}})\tanh ^{2} r}},
\end{eqnarray}
which are independence of the phase $\vartheta$. 

The corresponding form of $\mathcal{A}$ can be found as
\begin{eqnarray}
\mathcal{A}_{\left\vert \xi \right\rangle }=-\frac{i4\pi \exp({-8\pi \tilde{
\kappa}})\exp \left[ -i4\pi (\tilde{G}+2O)\right] \tanh ^{2} r}{1-\exp({-8\pi
\tilde{\kappa}})\exp \left[ -i4\pi \left( \tilde{G}+2O\right) \right] \tanh
^{2} r}.  \label{as}
\end{eqnarray}
Then applying Eqs.~(\ref{fish}) and (\ref{fish3}), we can acquire the QFI (see Tab.~1). 
Acorrding to Tab.~1, we introduce a ratio of the QFI    corresponding to the  squeezed vacuum state over that of  the coherent state at optimal points $O^{*}$  as
\begin{eqnarray}
\frac{F_{|\xi\rangle}}{F_{|\alpha\rangle}}= \frac{(3\bar{n}+2)\exp(4\pi\tilde{\kappa})}{[(\bar{n}+1)\exp(8\pi\tilde{\kappa})-\bar{n}]^2[1+\bar{n}\exp(-4\pi\tilde{\kappa})]}.
\end{eqnarray}
When  the ratio $F_{|\xi\rangle}/F_{|\alpha\rangle}>1$ ($F_{|\xi\rangle}/F_{|\alpha\rangle}<1$), it indicates that the squeezed states (coherent states) have advantage for quantum metrology. 
\begin{table*}[tbp]
\caption{ The  optimal
working points $O^{*}$, the optimal classical contribution $F_c^{*}$, the optimal quantum
contribution $F_q^{*}$ of the QFI, and the total QFI in the optimal working
points for coherent state $|\alpha\rangle$ and squeezed vacuum state $|\xi\rangle$. Here $\bar{n}=|\alpha|^2$ for coherent state $|\alpha\rangle$
 and $\bar{n}=\sinh^2 r$ for squeezed vacuum state $\xi\rangle$.}
\label{table1}
\begin{center}
\begin{tabular}{ccccc}
\hline\hline
 $|\phi\rangle$ & optimal points $O^{*}$ &  $F^{*}_c$ &  $F^{*}_q$ &  $F^{*}_{\kappa\to0}=F^{*}_c+F^{*}_q$ \\ \hline
$|\alpha\rangle$  & $\frac{m-\tilde{G}}{2}, m=0,\pm1,...$ & $16\pi^2\bar{n}e^{-4\pi\tilde{\kappa}}$ & $16\pi^2\bar{n}^2
e^{-8\pi\tilde{\kappa}}$ & $16\pi^2\bar{n}(\bar{n}+1) $ \\ \hline
$|\xi\rangle$  & $\frac{(m-2\tilde{G})}{4},m=0,\pm1,...$ & $\frac{32\pi^2 \bar{n}(\bar{n}+1)}{[(\bar{n}+1)e^{8\pi\tilde{\kappa}}-\bar{n}]^2}$
& $\frac{16\pi^2\bar{n}^2} {[(\bar{n}+1)e^{8\pi\tilde{\kappa}} - \bar{n}]^2}$
& $48\pi^2\bar{n}\left(\bar{n}+\frac{2}{3}\right)$ \\ \hline\hline
\end{tabular}
\end{center}
\end{table*}

\section{Supersensitive coupling-rate estimation \label{couprestimaiton}}
In this section, we discuss the Heisenberg-limited  estimation precision of the coupling rate in the presence of
photon loss, and give the optimal measurement scheme which can saturate the QCR bound.

\begin{figure}[ht]
\centering\includegraphics[scale=0.52]{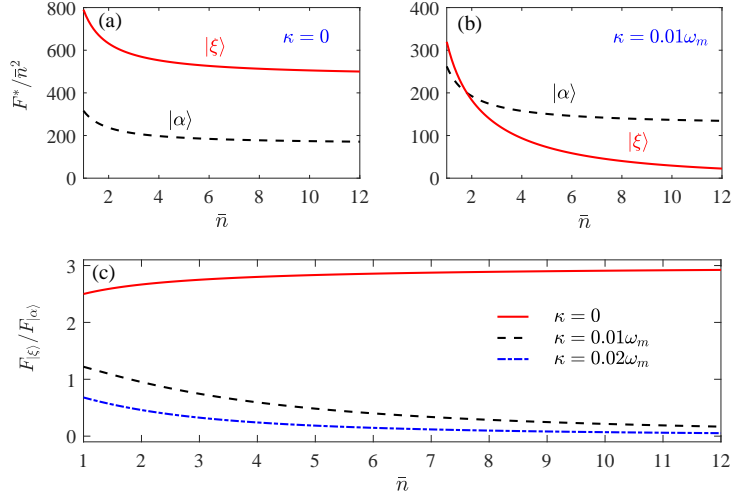}
\caption{The optimal rates of the Heisenberg limit $F^*/\bar{n}^2$  (a, b) and  $F_{\vert\xi\rangle}/F_{\vert\alpha\rangle}$  (c) with
respect to average number $\bar{n}$ for different values of photon loss $
\tilde{\kappa}=\kappa/\omega_{m}$.}
\end{figure}
To clearly see the behavior of the QFI with respect to $O$,
in Tab.~1 we compare the QFI corresponding to the cases of  coherent state and squeezed vacuum state.
As  it is shown, in our hybrid
system both these  input states can induce Heisenberg-limited QFI ($F \propto \bar{n}^2$),
which depends on $\tilde{G}$ and $O$. 
 For coherent state, the optimal values of QFI can be found
when $O\rightarrow 0.5(m-\tilde{G})$, $(m=0,\pm1,\pm2,...)$, while for squeezed vacuum 
state the optimal points are $O\rightarrow 0.5(0.5m-\tilde{G})$, $(m=0,\pm1,\pm2,...)$. 
 As expected, when $\tilde{\kappa}$ is not too large, as a
useful resource for quantum metrology, the squeezed vacuum state outperforms that of
coherent sate. 
  For instance,  when $\tilde{\kappa}\to 0$ and $\bar{n}= 1$, for squeezed vacuum state the optimal rate with respect to HL is $F_{|\xi\rangle}^*/\bar{n}^2\to 90\pi^2$,
which gains an advantage over that of the coherent state  case $F_{|\alpha\rangle}^*/\bar{n}^2 \to 32\pi^2$.
With the average photon number $\bar{n}$ increasing, both the rates will decrease and then tend to the steady values,
 $F_{|\xi\rangle}^*/\bar{n}^2=48\pi^2$ and $F_{|\alpha\rangle}^*/\bar{n}^2=16\pi^2$. What is more, we can also check that
the advantage of squeezed state for quantum
metrology in our system is attributed to the classical contribution $F_c$. This is because  $F^{|\xi\rangle}_c > F^{|\xi\rangle}_q\approx F^{|\alpha\rangle}_q>F^{|\alpha\rangle}_c$  when $\kappa \to 0$.

However, when increasing $\tilde{\kappa}$ the above results become
the  different.
In Fig.~2(a) and 2(b), we  plot the optimal rates of the HL $F^*/\bar{n}^{2}$ with respect to mean photon number $\bar{n}$ for different values of the loss rate $\tilde{\kappa}$, where  $F^*/\bar{n}^{2} \sim 1$  indicates the HL.
 As it is shown, when the photon loss rate $\tilde{\kappa}$ increasing, the advantage of squeezed state gradually loses.
 The same conclusion can be found in Fig.~2(c).
These results mean that for small  $\tilde{\kappa}$ and small mean photon number $\bar{n}$,
squeezed vacuum state shows advantage, but with the increase  of $\tilde{\kappa}$, the coherent state is more robust against the photon loss.

\begin{figure}[ht]
\centering\includegraphics[scale=0.54]{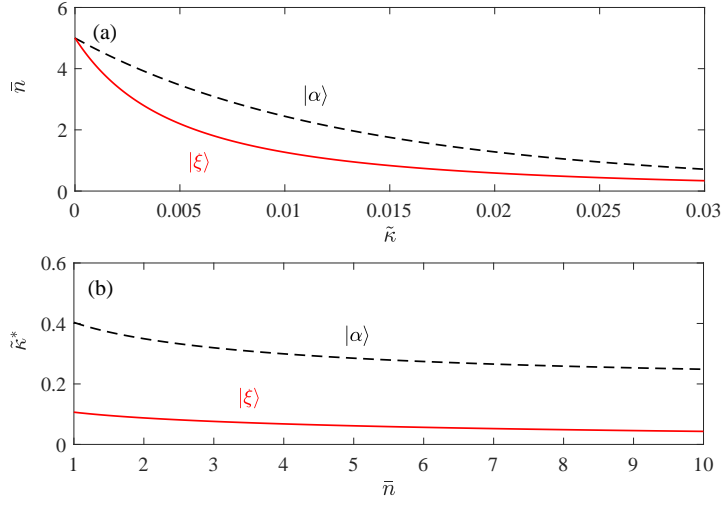}
\caption{ Part (a) shows the average photon number $\bar{n}_{|\alpha\rangle}$
and $\bar{n}_{|\xi\rangle}$ in the cavity as a function of loss rate $\tilde{\protect\kappa}$, where $\bar{n}=5$. Part (b) shows the
critical values $\tilde{\protect\kappa}^{*}$ of photon loss rate for
Heisenberg limit as a function of average photon number $\bar{n}$.}
\end{figure}

To understand  that coherent state can perform better for parameter estimation than  that of squeezed vacuum state in the presence of photon loss,
below we will investigate the influence of the cavity photon loss on the estimation precision.
As the QFI depends on average cavity photon number, the decrease of the average cavity photon number will result in the estimation sensitivity reduce.
Thus, we first calculate the average cavity photon number for the above input states with photon loss and then give the critical conditions of HL for them.
At time $\tau =2\pi/\omega _{m}$, by trace over the impurity part and the Bogoliubov mode from the total density matrix, the matrix
elements of the cavity optical modes read
\begin{align}
\rho_{mn}(\tau) =  & \hspace{0.1cm}\frac{1}{2} \exp[-2\pi \tilde{\kappa} (m+n)] \exp \{-i2\pi  [\tilde{\Omega}_{c}(m-n) - \tilde{\chi}^{2}(m^{2}-n^{2})] \} \nonumber\\
&\times \{ \exp[-i2\pi(\tilde{G}+2O)(m-n)] +1 \} \langle m |\phi\rangle \langle \phi| n\rangle. 
\end{align}
Then the average photon number $\bar{n}$ in the cavity for coherent state and squeezed vacuum state  can be respectively  obtained as
\begin{eqnarray}
&&\bar{n}_{\vert\alpha\rangle}=\sum_{n}n\rho_{nn}=|\alpha|^{2}\exp\left[|\alpha|^{2}(e^{-4\pi\tilde{\kappa}}-1)-4\pi\tilde{\kappa}\right],\notag\\
&&\bar{n}_{\vert\xi\rangle} =\sum_{n}2n\rho_{2n2n}=\frac{\sech r\tanh^{2} r\exp({4\pi\tilde{\kappa}})}{\left[\exp({8\pi\tilde{\kappa}})-\tanh^{2} r\right]^{3/2}}.
\end{eqnarray}
Figure~3(a) shows the average photon number $\bar{n}$ in the cavity as a
function of loss rate $\tilde{\kappa}$. It indicates that with the increase  of $\tilde{\kappa
}$, $\bar{n}_{|\xi \rangle }$ demonstrates  faster decay than that of $\bar{n}_{|\alpha \rangle }$.

 To maintain the Heisenberg-limited sensitivity, $F\sim \bar{n}^{2}$, the values of photon loss
$\tilde{\kappa}$ should be below a critical condition. By calculating equation $F(\tilde{\kappa}^{\ast })=\bar{n}^{2}$, we can obtain the critical values for coherent state
and squeezed vacuum state, respectively, as
\begin{eqnarray}\label{gamac}
\tilde{\kappa}_{|\alpha \rangle }^{\ast } =\frac{1}{4\pi }\ln \left[ \frac{
4\pi \bar{n}}{\sqrt{4\pi ^{2}+\bar{n}^{2}}-2\pi }\right] , \hspace{0.6cm} \tilde{\kappa}_{|\xi \rangle }^{\ast } =\frac{1}{8\pi }\ln \left[ \frac{\bar{n}^{2}+4\pi \sqrt{\bar{n}(3\bar{n}+2)}}{\bar{n}(\bar{n}+1)}\right] .
\end{eqnarray}
 Based on Eq.~(\ref{gamac}), in Fig.~3(b) we  plot the
critical values $\tilde{\kappa}^{\ast }$ of photon loss rate for HL as a function of average photon number $\bar{n}$. We can
see from Fig.~3(b) that the critical values $\tilde{\kappa}^{\ast }$ for
coherent state is always larger than that of the squeezed vacuum state,
which indicates that the coherent state is more robust against the photon loss.
These results can be used to explain the phenomena shown in Fig.~2.
\begin{figure}[ht]
\centering\includegraphics[scale=0.54]{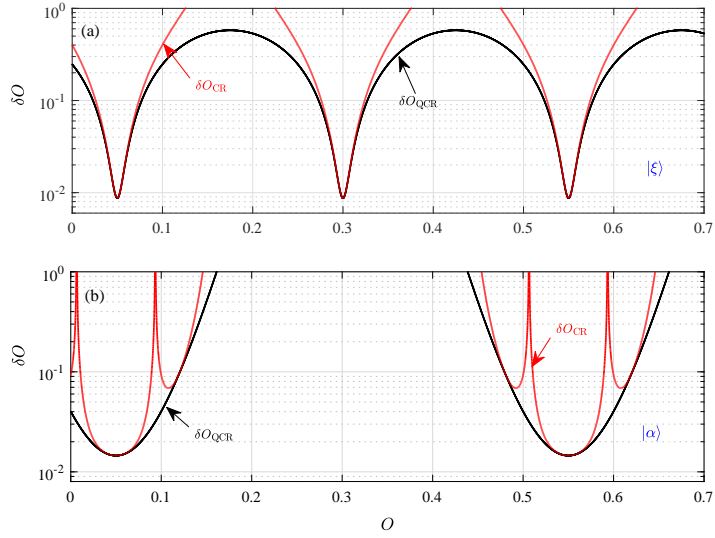}
\caption{The precision $\delta O$ on the estimated parameter $O$ for
squeezed vacuum state (a) and coherent state (b).  Here 
 $\delta O_{\mathrm{QCR}}\equiv 1/\protect
\sqrt{F}$ and $\delta O_{\mathrm{CR}}\equiv 1/\protect
\sqrt{F_{\rm cl}}$. 
Other parameters are $\bar{n}=5$, $\tilde{G}=-0.1$, and
$\kappa \to 0$.}
\end{figure}

In practical experiments, to find the optimal measurement that can lead to
the highest precision of the estimated parameter is of practical importance.
The classical Fisher information (CFI) can determine the minimum
standard deviation of a parameter estimator once we have chosen
a single specific measurement   with POVM elements.
Here, we consider a fixed measurement $\{|+\rangle\langle+|,|-\rangle\langle-|\}$ 
[here $|\pm\rangle=(|e\rangle\pm|g\rangle)/\sqrt{2}$] to
extract the information of the estimated parameter $O$ \cite{liu3}, which
can be realized by using a
very fast $\pi/2$ pulse to the impurity atom followed a projective measurement   $\{|e\rangle\langle e|, |g\rangle\langle g| \}$ 
 of the atomic states.  This $\pi/2$ pulse is finished in a very short duration of time and hence the evolution of other systems is neglected during this period. In this sense, 
 the backaction of the measurement of the impurity atom can be approximately ignored.
Therefore,  the CFI is given by
\begin{eqnarray}
F_{\mathrm{cl}}=\sum_{\pm}\frac{[\partial_O P(\pm|O)]^2}{P(\pm|O)},
\end{eqnarray}
where $P(\pm|O)=\langle\pm|\rho|\pm\rangle$ is the probability of getting
the measurement result at the eigenvalues of $|\pm\rangle$. Through some
calculations, we have
$P(\pm|O)=\frac{1}{2}\pm \frac {\mathrm{Re}[\rho_{eg}]}{2\rho_{ee}}$,
and the CFI can be obtained as
\begin{eqnarray}
F_{\mathrm{cl}}=\frac{(\text{Re}[\mathcal{A}] \text{Re}[\rho_{eg}]-\text{Im}[\mathcal{A}]
\text{Im}[\rho_{eg}])^2}{\rho_{ee}^{2}-(\text{Re}[\rho_{eg}])^{2}}.
\end{eqnarray}
Then the standard deviation for the parameter $O$ is  given by the Cram\'er-Rao (CR) bound
\begin{eqnarray}  \label{deo}
\delta O_{\mathrm{CR}} \equiv \frac{1}{\sqrt{F_{\mathrm{cl}}}}=\frac{\sqrt{\rho_{ee}^{2}-(\text{Re}[\rho_{eg}])^{2}}}{\vert \text{Re}[\mathcal{A}]\text{Re}[\rho_{eg}]-\text{Im}[\mathcal{A}]\text{Im}[\rho_{eg}]\vert},
\end{eqnarray}
usual we have $\delta O_{\mathrm{CR}}\geq\delta O_{\mathrm{QCR}}$.

Note that the precision obtained by the fixed measurement $\{|+\rangle\langle+|,|-\rangle\langle-|\}$ is equivalent to the measurement of the
observable Pauli operator $\sigma_{x}$.
The expectation of the measurement is given as
\begin{eqnarray}
\langle\sigma_{x}\rangle=\frac{\rho_{eg}+\rho_{eg}^{\ast}}{2\rho_{ee}}=\frac{\text{Re}[\rho_{eg}]}{\rho_{ee}},
\end{eqnarray}
which is a periodic signal of $O$.
Then we arrive at
\begin{eqnarray}
\frac{d\langle\sigma_{x}\rangle}{dO}=\frac{\text{Re}[\mathcal{A}]\text{Re}[\rho_{eg}]-\text{Im}
[\mathcal{A}]\text{Im}[\rho_{eg}]}{\rho_{ee}},
\end{eqnarray}
which reflects the sensitivity of measurement signal to parameter $O$, and
the estimation fluctuation is $\Delta\sigma_{x}=\sqrt{1-\langle\sigma_{x}\rangle^{2}}$.
Based on the error propagation formula, the minimum standard deviation for the
coefficient $O$ is given by
\begin{eqnarray}
\delta O_{\rm CR}=\frac{\Delta\sigma_{x}}{\vert d\langle\sigma_{x}\rangle/dO\vert}=\frac{1}{\sqrt{F_{\rm cl}}},
\end{eqnarray}
which is the same as Eq.~(\ref{deo}). The highest sensitivities can be found at points $O^*$ which have been given in Tab.~1.
By calculating Eqs.~(\ref{deo}) and ~(\ref{fish}), as shown in Fig.~4, we find that the
coupling strength sensitivity $\delta O_{\rm CR}$ can be in accordance with the best
sensitivity of $\delta O_{\rm QCR}$ at  points $O^*$.

\section{Conclusion \label{conclusion}}

We have presented a  scheme to obtain HL coupling-rate estimation
in a hybrid cavity optomechanics formed by a cavity-BEC system coupled to an impurity.
In our scheme we use the impurity atom as a probe qubit to detect the coupling-rate by encoding it into the dynamical phase of the impurity atom.
Through  calculating the QFI of the impurity, we found that the parameter sensitivity can be remarkably
improved  by using either coherent state or squeezed vacuum state fed into the cavity. For small
 photon decay rate, compared with the case of coherent state the squeezed
vacuum state has some advantage owing to the classical contribution of the
QFI. We also found that the HL can maintain even in the presence of
cavity photon loss. Through comparing the critical values of photon loss
rate, we found that the coherent state is more robust against the photon loss than squeezed vacuum state.
Furthermore, we have demonstrated that the superprecision given by the
QCR bound can be realized by implementing the fixed measurement
$\{|+\rangle\langle+|,|-\rangle\langle-|\}$ on the impurity atom. 

We would like to remark that the squeezed coherent state may have the advantages of both squeezed vacuum and coherent states, 
and it is worth  study in details.
Finally, we hope that our study might have promising
applications in high-precision estimations of the coupling strength that may carry an important quantity in  a hybrid optomechanical system.

\appendix*

\section{Derivation of the effective Hamiltonian in the large-detuning case}
In this Appendix, we present a detailed derivation of the effective Hamiltonian of the system in
the large-detuning case.
In the frame rotating at the cavity frequency $ \omega_{c}$, the many-body Hamiltonian of the quasi-1D BEC reads~\cite{Maschler, Zhang, Yang}
\begin{eqnarray}
H_{\text{BEC}} &=&\int dx\left[ {\Psi}_{g}^{\dag }(x)\hat{h}(x) {\Psi}_{g}(x) +{\Psi}_{e}^{\dag}(x)\left( \hat{h}(x)+\hbar \Delta_{ac}\right) {\Psi}_{e}(x)\right]\nonumber\\
&&+\frac{g_{1D}}{2}\int dx{\Psi}_{g}^{\dag }(x){\Psi}_{g}^{\dag }(x){\Psi}_{g}(x){\Psi}_{g}(x), 
\end{eqnarray}
where $\hat{h}(x)= -\frac{\hbar^{2}}{2m}\frac{d^{2}}{dx^{2}} +V_{\rm vex}(x)$ is the single-particle Hamiltonian, and
 $\Delta_{ac}=\omega_{a}-\omega_{c}$ is the detuning between the BEC atoms and the cavity field, $\Psi_{g}(x)$ and $\Psi_{e}(x)$ denote the atomic field operators, which describe the annihilation of an atom at position $x$ in the ground and excited states, respectively. They obey the usual
bosonic commutation relations
\begin{eqnarray}
\lbrack {\Psi}_{f}(x),{\Psi}_{f^{\prime }}^{\dag }(x^{\prime })]
=\delta (x-x^{\prime })\delta _{ff^{\prime }}, \hspace{0.5cm}
\lbrack {\Psi}_{f}(x),{\Psi}_{f^{\prime }}(x^{\prime })] = [{
\Psi}_{f}^{\dag }(x),{\Psi}_{f^{\prime }}^{\dag }(x^{\prime })]=0
\end{eqnarray}
for $f,f^{\prime }\in \{e,g\}$. 
Here, $V_{\rm ext}(x)$ denotes the weak external harmonic trapping. 

The interaction between the BEC and the cavity is described by the Hamiltonian
\begin{equation}
H_{\text{cav-BEC}}^{\rm int}=\hbar g\cos (kx)\int dx\left[ {\Psi}_{g}^{\dag}(x)a^{\dag }{\Psi}_{e}(x)+{\Psi}_{e}^{\dag }(x)a{\Psi}_{g}(x)\right].
\end{equation}
In this rotating frame, the Hamiltonian of the two-level impurity atom reads $H_{\text{imp}}=\hbar \Delta
\left\vert e\right\rangle \left\langle e\right\vert$. 
The Hamiltonian describing the
cavity-impurity coupling is written as
\begin{equation}
H_{\mathrm{cav-imp}}=\hbar g_{ac}(a^{\dag }\sigma^{-} _{a}+a\sigma ^{+}_{a}),
\end{equation}
where  $\sigma_{a}^{+}=(\sigma_{a}^{-})^{\dagger}=|e\rangle\langle g|$ are
the ladder operators for the impurity, and $g_{ac}$  is the
maximum impurity-cavity coupling strength.

We assume that the BEC atoms are only coupled with the ground state $\left\vert g\right\rangle$ of the impurity atom via the Raman transition
\begin{equation}
H_{\text{BEC-imp}}=\hbar \bar{g}_{\text{BEC-imp}}^{1D}\left\vert g\right\rangle \left\langle g\right\vert \int dx\left\vert \varphi
\right\vert ^{2}{\Psi}_{g}^{\dag }(x){\Psi}_{g}(x),
\end{equation}
where $\bar{g}_{\text{BEC-imp}}^{1D}$ is the 1D interaction strength which
can be obtained by integrating out the $y$ and $z$ variables of the atomic
field operators
\begin{eqnarray}
\bar{g}_{\text{BEC-imp}}^{1D} =\frac{4\pi \hbar a_{AB}}{m_{AB}}\int
dydz\left\vert \varphi _{\bot }(y,z)\right\vert ^{2}\Psi _{\bot }^{\dag
}(y,z)\Psi _{\bot }(y,z)
=\frac{4\hbar a_{AB}}{m_{AB}(\ell _{A\bot }^{2}+\ell _{B\bot }^{2})},
\end{eqnarray}
where  $\Psi_{\bot}(y,z)=(\pi \ell_B^2)^{-1/2}e^{-(y^2+z^2)/(2\ell_B^2)}$
and $\varphi_{\bot}(y,z)=(\pi \ell_A^2)^{-1/2}e^{-(y^2+z^2)/(2\ell_A^2)}$.

Based on the above analyses, the Hamiltonian of the whole system can be expressed as
\begin{eqnarray}
H=H_{\mathrm{BEC}}+H_{\mathrm{cav-BEC}}^{\rm int}+H_{\mathrm{imp}}+H_{\mathrm{cav-imp}}+H_{\mathrm{BEC-imp}}.
\end{eqnarray}
Accordingly, the Heisenberg equations for the various field
operators are obtained as
\begin{eqnarray}
\frac{\partial {\Psi}_{e}(x)}{\partial t} =-\frac{i}{\hbar }[{\Psi}_{e}(x),H]
=i\left( \frac{-\hbar ^{2}}{2m}\frac{d^{2}}{dx^{2}}-\Delta _{ac}\right)
{\Psi}_{e}(x)-ig(x)a {\Psi}_{g}(x),
\end{eqnarray}
and
\begin{eqnarray}
\frac{\partial \sigma _{a}^{+}}{\partial t} =-\frac{i}{\hbar }[\sigma_{a}^{+},H]
=i\Delta \sigma _{a}^{+}-i\sigma _{a}^{+}\bar{g}_{\text{BEC-imp}}^{1D}\int dx\left\vert \varphi
\right\vert ^{2}{\Psi}_{g}^{\dag }(x){\Psi}_{g}(x)-ig_{ac}\sigma_{a}^{z}a^{\dag}.
\end{eqnarray}

Below, we assume that the temperature of the BEC is sufficient low to neglect the thermal excitation. In particular, we consider the case where both the BEC-atom-cavity coupling and the impurity-cavity coupling work in the large-detuning regime, and then the quantum coherence transition  of the impurity and the BEC atoms can be eliminated adiabatically. To this end, we  neglect the kinetic energy term and  trapping potential of the BEC atoms in the zero-temperature limit, because it is much smaller than the internal energy term~\cite{Maschler}. Then the excited-state atomic field operator and the transition operator of the impurity atom can be expressed as
\begin{eqnarray}
{\Psi}_{e}(x)=-\frac{g(x)a{\Psi}_{g}(x)}{\Delta _{ac}},  \hspace{0.5cm}
\sigma _{a}^{+}=\frac{g_{ac}a^{\dag }\sigma _{a}^{z}}{\Delta -\bar{g}_{\text{BEC-imp}}^{1D} \int
dx\vert \varphi\vert^{2}{\Psi}_{g}^{\dag }(x){\Psi}_{g}(x)}
\approx \frac{g_{ac}a^{\dag }\sigma _{a}^{z}}{\Delta },
\end{eqnarray}
where we have made the approximation  based on the fact that $\Delta\gg \bar{g}_{\text{
BEC-imp}}^{1D}$~\cite{yuan,song}.
Inserting the above equations for ${\Psi}_{e}(x)$ and $\sigma _{a}^{+}$
into $H$, we arrive at
\begin{eqnarray}
H_{\rm eff} &=&\int dx{\Psi}_{g}^{\dag }(x)\left( \frac{-\hbar ^{2}}{2m}\frac{d^{2}
}{dx^{2}}+V_{\rm ext}(x)+\hbar U_{0}\cos ^{2}(kx)a^{\dag }a+\frac{g_{1D}}{2}\left\vert {
\Psi}_{g}(x)\right\vert ^{2}\right) {\Psi}_{g}(x) \nonumber\\
&&+\hbar \Delta \vert e\rangle \langle
e\vert+\hbar \bar{g}_{\text{BEC-imp}}^{1D}\left\vert g\right\rangle \left\langle
g\right\vert \int dx\vert \varphi \vert ^{2}{\Psi}_{g}^{\dag
}(x){\Psi}_{g}(x) +\delta_{ac}\sigma_{a}^{z}(2 a^{\dag }a+1),
\end{eqnarray}
where $U_{0}=-{g^{2}}/{\Delta _{ac}}$ and $\delta_{ac}=g^2_{ac}/{\Delta}$.
In the main text, we denote ${\Psi}_{g}(x)$ as ${\Psi}(x)$ for concise.

\section*{Funding}

Natural National Science Foundation of china (11805047, 11665010,11822501,111775075,1905053); 
Natural Science Foundation of Hunan Province(2018JJ3006, 2017JJ1021); Open Fund Project of
the Hunan Provincial Applied Basic Research Base of Optoelectronic Information Technology(GD19K05);
Hainan Association for Science and Technology (QCXM201810).

Hunan Provincial Natural Science Foundation of China (2017JJ1021); Hunan Provincial Natural Science Foundation of China under Grant No2018JJ3006 and  the Open Fund Project of the Hunan Provincial Applied Basic Research Base of Optoelectronic Information Technology under Grant No GD19K05.
Young Talents'Science and Technology Innovation Project of Hainan Association for Science and Technology (QCXM201810).
\section*{Acknowledgments}
Q.S.T. thanks Professor J. Liu for valuable discussion.
 Q.S.T  acknowledges  from  the National Natural Science Foundation of China (NSFC) (11665010,11805047).
Y.J.B  is Supported by the NSFC (11905053) and Hunan Provincial Natural Science Foundation of China under Grant No2018JJ3006 and  the Open Fund Project of the Hunan Provincial Applied Basic Research Base of Optoelectronic Information Technology under Grant No GD19K05.
J.-Q.L. is supported in part by National Natural Science Foundation of China (Grants No. 11822501, No. 11774087, and No. 11935006) and Hunan Science and Technology Plan Project (Grant No. 2017XK2018). L.M.K is supported by the NSFC (11775075, 11434011).

\section*{Disclosures}
The authors declare no conflicts of interest.

\end{document}